\documentclass[aps,pre,amsmath,twocolumn,superscriptaddress]{revtex4-1}
\usepackage[figuresright]{rotating}
\usepackage{braket}
\usepackage{bm}
\usepackage{amsmath,amssymb}
\usepackage{color}
\usepackage{url}
\usepackage{algorithm}
\usepackage{algpseudocode}
\usepackage[titletoc]{appendix}
\usepackage{bbm}
\usepackage{bbold}
\usepackage{mathrsfs}

\usepackage[utf8]{inputenc}

\newcommand{\im}{{i}}          

\begin{document}

	\title{Quantum Neimark-Sacker bifurcation}
\author{I.I.~Yusipov and M.V.~Ivanchenko}

\affiliation{Department of Applied Mathematics, Lobachevsky University, Nizhny Novgorod, Russia}

\begin{abstract} 
	Recently, it has been demonstrated that asymptotic states of open quantum system can undergo qualitative changes resembling pitchfork,  saddle-node, and period doubling classical bifurcations. Here, making use of the  periodically modulated open quantum dimer model, we report and investigate a quantum Neimark-Sacker bifurcation. Its classical counterpart is the birth of a torus (an invariant curve in the Poincar\'{e} section) due to instability of a limit cycle (fixed point of the Poincar\'{e} map). The quantum system exhibits a transition from unimodal to bagel shaped stroboscopic distributions, as for Husimi representation, as for observables. The spectral properties of Floquet map experience changes reminiscent of the classical case, a pair of complex conjugated eigenvalues approaching a unit circle. Quantum Monte-Carlo wave function unraveling of the Lindblad master equation yields dynamics of single trajectories on ``quantum torus'' and allows for quantifying it by rotation number. The bifurcation is sensitive to the number of quantum particles that can also be regarded as a control parameter.  
	
\end{abstract}

\maketitle

\section{Introduction}\label{sec:1}

Bifurcation analysis, introduced by Poincar\'{e} more than a century ago \cite{Poincare1885}, has become a primary approach in nonlinear dynamics and applications \cite{Kuznetsov,DE}. Its extensions to the quantum realm followed much later. The first and most celebrated example of bridging complex dynamics in the two domains is Hamiltonian chaos, which spectral signatures in quantum systems are profoundly understood by now  \cite{Casati1979,Gutzwiler1991,Haake1992,Guhr1998}. For quite a while, the studies kept focused on Hamiltonian systems, seeking the footprints of bifurcations in the classical phase space on the properties of corresponding quantum equations and their solutions. The archetypal pitchfork and Hopf bifurcations \cite{Kuznetsov,DE} in the mean-field equations were related to sharp changes of the ground-state entanglement in the corresponding quantum models \cite{bif0,bif2}. A pitchfork bifurcation was also found to underpin the transition from Rabi to Josephson dynamics in experiments with rubidium Bose-Einstein condensate~\cite{bif3}. 

Recent experimental advances in cavity quantum electrodynamics \cite{Walter2006}, quantum optical systems \cite{Aspelmeyer2014}, artificial atoms \cite{You2011}, polaritonic devices \cite{Feurer2003}, and superconducting circuits \cite{Barends2015,Roushan2017} have spurred attention to open quantum systems. These systems interact with their environments (or are subjected to actions from outside), and therefore their dynamics is essentially 
dissipative \cite{Carmichael1991,book}. Although such systems evolve to a unique asymptotic state under broad conditions \cite{spohn,alicki}, the dynamics has proved to be no less complex then the unitary one, framed by a set of eigenstates, yielding the states structurally and dynamically similar to classical chaotic attractors \cite{Spiller1994,Brun1996,Carlo2005,Hartmann2017,Ivanchenko2017,Poletti2017,Yusipov2019}.   

To date, several quantum counterparts of dissipative bifurcations have been described: pitchfork, saddle-node, and period doubling \cite{bg3,bg2,Hartmann2017,Ivanchenko2017,Poletti2017,Poletti2018}. This is  commonly done in the Markovian framework, and the dynamics of a model system is described with Lindblad equation \cite{lind,gorini, alicki,book}. The nonlinear mean-field equations used as a classical reference are obtained following Bogoliubov-Born-Green-Kirkwood-Yvon (BBGKY)-like approaches, by truncating the hierarchy of cumulants on the level of expectation values \cite{bg3} or keeping double correlators \cite{bg1,bg2}. Routinely, quantum bifurcations are visualized by calculating quasi-classical phase space distributions, Husimi or Wigner-like \cite{stock}, which structural changes with a bifurcation parameter reproduce bifurcations in the classical phase space. For instance, the quantum period-doubling bifurcation is seen as the transition from unimodal to bimodal Husimi distribution \cite{Hartmann2017,Poletti2018}. At the same time, it has been demonstrated that quantum bifurcations can be observed directly as the structural changes in the asymptotic density matrix \cite{Ivanchenko2017}. This approach allows to overcome the technical limitation of calculating Husimi distribution for the system size $N>10^3$, when it becomes computationally unfeasible. Both approaches, however, do not resolve dynamics {\it on the quantum attractor}. 

In this paper we find and study the quantum Neimark-Sacker bifurcation, which classical counterpart is the birth of a torus (an invariant curve in the Poincar\'{e}/stroboscopic section) due to instability of a limit cycle (fixed point of the Poincar\'{e} map)  \cite{Kuznetsov}. Exemplifying in the experimentally relevant open quantum periodically modulated dimer model, we show that the stroboscopic Husimi distribution exhibits a transition from the unimodal to bagel-shaped form -- a section of ``quantum torus'' -- for the boson interaction strength close to the bifurcation value for the mean-field model. Importantly, the same transformation is observed in the stroboscopic distribution of observables, obtained by the Monte Carlo wave-function stochastic unraveling of the Lindblad equation, the  method, especially relevant in the context of quantum optics and cavity systems \cite{zoller,dali,plenio,daley}. Dynamics of such individual quantum trajectories  on ``quntum torus'' is suitably characterized by rotation number. Similar to the classical case, rotation numbers close to rational correspond to almost ``periodic'' multi-modal stroboscopic distributions. Finally, we demonstrate that the bifurcation is system size dependent, which plays a role of another bifurcation parameter.

The paper is organized as follows. In Section \ref{sec:3} we describe the quantum model, its nonlinear mean-field approximation, and numerical methods. Section \ref{sec:7} contains the main results. Section \ref{sec:8} gives a summary and outlook.

\section{Model and Methods} \label{sec:3}

Within the Markovian approximation framework (which assumes weak coupling to environment), the evolution of an open quantum system can be described by the Lindblad master equation \cite{book, alicki},
\begin{align}
\dot{\varrho} = \mathcal{L}(\varrho) = -\im [H,\varrho] + \mathcal{D}(\varrho),
\label{eq:1}
\end{align}
where the first term in the r.h.s.\ captures the unitary evolution, 
and the second term describes the action of environment. We consider a system of $N$ indistinguishable interacting bosons, 
that hop between the sites of a periodically rocked dimer. This model is  a popular theoretical testbed  \cite{vardi, trimborn, poletti}, 
recently implemented in experiments \cite{weiss,oberthaler, ober1}, 
known to exhibit regular and chaotic regimes \cite{Hartmann2017,Ivanchenko2017,Poletti2017,Poletti2018}. Its unitary dynamics is governed by the Hamiltonian 
\begin{align}
H(t)=&-J \left( b_1^{\dagger} b_2 + b_2^{\dagger} b_1 \right) + \frac{2U}{N} \sum_{g=1,2} n_g\left(n_g-1\right)\nonumber \\
&+\varepsilon(t)\left(n_2 - n_1\right) \;. 
\label{eq:2}
\end{align}
Here, $J$ denotes the tunneling amplitude, $U$ is the interaction strength, and $\varepsilon(t)$ presents a periodical modulation of the on-site potentials. It is chosen as $\varepsilon(t) = \varepsilon(t + T) = A\sin(\Omega t)$,  $\Omega=2\pi/T$, so that the amplitude $A$ is the dynamic energy offset between the two sites. Here $b_j$ and $b_j^\dag$ are the annihilation and creation operators of a particle at site $j$, while $n_j = b_j^\dag b_j$. 

The system Hilbert space has dimension $N+1$ and can be spanned  with $N+1$ Fock basis vectors, labeled by the number of bosons on the first  site, $\{|n\rangle\}$, $n=0,...,N$. Thus, the size of the model is controlled by the total number of bosons.

The dissipative term involves a single experimentally relevant jump operator \cite{Diehl2008,kraus,Diehl2013}:
\begin{align}
\mathcal{D}(\varrho) =& \frac{\gamma}{N} \left(V\varrho V^\dagger - \frac{1}{2}\{V^\dagger V,\varrho\}\right), \\
V=&(b_1^{\dagger} + b_2^{\dagger})(b_1-b_2), 
\label{eq:3}
\end{align}
which attempts to `synchronize' the dynamics on the two sites by constantly recycling anti-symmetric out-phase modes into symmetric in-phase ones. The dissipative coupling constant $\gamma$ is taken  to be
time-independent.

Throughout the paper we set $J=1$, $\gamma=0.1$, $A=3.4$, $T=2\pi$, and vary $U$ and $N$.

The computational analysis makes use of two methods to evolve the system numerically. 
First, we implement propagation of Eq.(1) by the fourth-order Runge-Kutta scheme. The evolution converges to a unique asymptotic solution, which in case of periodic modulation is a stable periodic trajectory (or a fixed point of the corresponding stroboscopic map, $\mathcal{P}_F~:~\rho(mT)\rightarrow\rho((m+1)T)$, $m=0, 1, 2, \ldots$). For the particular system Eqs.(\ref{eq:2}),(\ref{eq:3}), we set the integration time step of $5\cdot10^{-4}T$, and leave at least $100$ modulation periods for transients to fade out. 

The density matrix $\varrho$ of the system with $N$ bosons can be visualized on the Bloch sphere by plotting the Husimi distribution $p\left(\vartheta, \varphi\right)$, which can be obtained by projecting $\varrho$ on the set of the generalized SU(2) coherent states \cite{arecchi, perelomov}
\begin{equation}\label{eq:3_1}
|\vartheta , \varphi \rangle = \sum_{j=0}^{N}\sqrt{\binom{N}{j}}\left[\cos{\frac{\vartheta}{2}}\right]^j \left[e^{i\phi}\sin{\frac{\vartheta}{2}}\right]^{N-j} |j\rangle.
\end{equation}

 The corresponding nonlinear mean-field equations for the $(\theta,\phi)$ phase variables are obtained by writing the master equation in terms of the spin operators $\mathscr{S}_x = \frac{1}{2N} \left(b_1^\dagger b_2 + b_2^\dagger b_1\right)$, $\mathscr{S}_y = -\frac{i}{2N} \left(b_1^\dagger b_2 - b_2^\dagger b_1\right)$, $\mathscr{S}_z = \frac{1}{2N} \left(n_1 - n_2\right)$, and considering their evolution in the Heisenberg picture \cite{book,Hartmann2017,Ivanchenko2017}. For a large number of $\left(N-1\right)$ bosons, the commutator $\left[\mathscr{S}_x, \mathscr{S}_y\right] = i \mathscr{S}_z \stackrel{N\rightarrow\infty}{=} 0$ vanishes, as well as the similar ones for other cyclic permutations. Replacing operators with their expected values, $\mathscr{S}_k = tr\left[\varrho\mathscr{S}_k\right]$, and denote $\mathscr{S}_k$ by $S_k$, one can obtain

\begin{eqnarray}
\begin{matrix}
\frac{dS_x}{dt} = 2\varepsilon\left(t\right) S_y - 8US_z S_y + 8\gamma\left(S_y^2+S_z^2\right), \\
\frac{dS_y}{dt} = -2\varepsilon\left(t\right) S_x + 8US_x S_z - 2JS_z + 8\gamma S_x S_y, \\
\frac{dS_z}{dt} = -2JS_y - 8\gamma S_x S_z, 
\end{matrix}
\label{eq:3_2}
\end{eqnarray}
where proportional to $\gamma$ terms of lower order in $N$ are neglected. The quantity $S^2 = S_x^2 + S_y^2 + S_z^2$ is a constant of motion, so the mean-field evolution is restricted to the surface of a Bloch sphere, $\left(S_x, S_y, S_z\right) = \frac{1}{2}\left[\cos{\left(\varphi\right)}\sin{\left(\vartheta\right)}, \sin{\left(\varphi\right)}\sin{\left(\vartheta\right)}, \cos{\left(\vartheta\right)}\right]$, yielding the equations of motion

\begin{eqnarray}
\begin{matrix}
\dot{\vartheta} = -2J\sin{\left(\varphi\right)} + 4\gamma \cos{\left(\varphi\right)}\cos{\left(\vartheta\right)}, \\
\dot{\varphi} = -2J\frac{\cos{\left(\vartheta\right)}}{\sin{\left(\vartheta\right)}} - 2\varepsilon\left(t\right) + 4U \cos{\left(\vartheta\right)} - 4\gamma\frac{\sin{\left(\varphi\right)}}{\sin{\left(\vartheta\right)}}. 
\end{matrix}
\label{eq:3_3}
\end{eqnarray}
The corresponding particle number in the first site is then recovered as $n=\frac{N}{2}(1+\cos{\left(\vartheta\right)})$. 

This nonlinear dynamical system plays a reference role in the bifurcation analysis further. In numerical experiments we record the consecutive stroboscopic values, $\{\vartheta(mT),\varphi(mT)\}, m=0, 1, 2, \ldots$, and construct the histograms normalized to the largest value $1$ for each parameter set, complemented by Poincar\'{e} sections for the stroboscopic map.  

Second, we employ the Monte-Carlo wave function (also ``quantum trajectory'' or ``quantum jump'') method \cite{zoller,dali}	to unravel  Lindblad master equation (\ref{eq:1}) into an ensemble of quantum trajectories. It recasts the evolution of the model system into the ensemble of systems described by wave functions, $\psi_r(t)$, $r = 1,2,...,M_r$,  
governed by an effective non-Hermitian Hamiltonian, $\tilde{H}$. This Hamiltonian incorporates the dissipative operator $V$, which is responsible for the decay of the norm,
\begin{align}
i\dot{\psi}=\tilde{H}\psi, \ \tilde{H} = H -\frac{i}{2} V^\dagger V.
\label{eq:4a}
\end{align} 
When the norm drops below a threshold, that is randomly chosen each time, a quantum jump is simulated, such that the wave function is transformed according to $\psi\rightarrow V \psi$ and then normalized \cite{Carmichael1991}.

The density matrix can then be sampled from a set of $M_r$ realizations as
$\varrho(t_\mathrm{p};M_{\mathrm{r}}) = \frac{1}{M_{r}}
\sum_{j=1}^{M_{\mathrm{r}}} \ket{\psi_j(t_\mathrm{p})}\bra{\psi_j(t_\mathrm{p})}$, which, given an initial pure state $\psi^\mathrm{init}$, 
converges towards the solution of Eq.~(\ref{eq:1}) at time $t_\mathrm{p}$ for the initial density matrix $\varrho^{\mathrm{init}} = \ket{\psi^\mathrm{init}}\bra{\psi^\mathrm{init}}$. 
We make use of the recently developed high-performance realization of 
the method \cite{Volokitin2017} and generate $M_{\mathrm{r}}=10^2$ different trajectories for averaging, 
leaving $t_0 = 2\cdot10^3 T$ time for relaxation towards an asymptotic state, and following the dynamics for up to $t = 10^3T$.

Thus we produce stroboscopic plots for the expectation values of the two observables, the number of particles on the left site of the dimer, $n(t)$, and the energy, $e(t)$,  
\begin{align}
n(t)&= \langle\psi(t)|b^\dagger_1b_1|\psi(t)\rangle, \\
e(t)&=\langle\psi(t)|H|\psi(t)\rangle.
\label{eq:5a}
\end{align}
Quantum trajectories allow for an insight to dynamics in the asymptotic regime -- on {\it quantum attractor} -- beyond the stationary Husimi projection picture.

\section{Results}\label{sec:7}

\begin{figure}
	\includegraphics[width=0.5\textwidth]{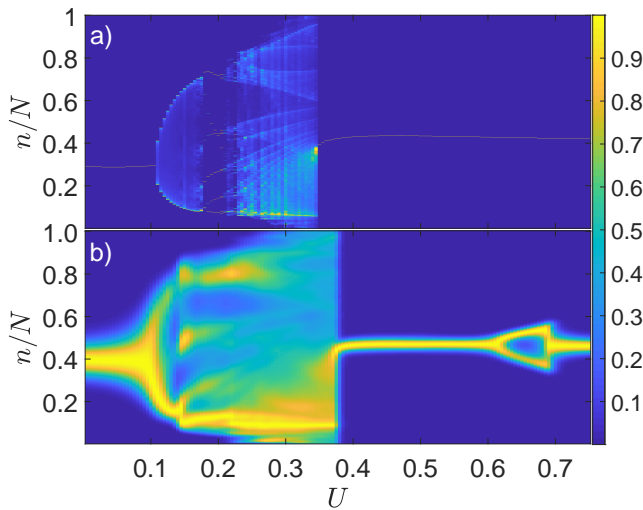}
	\caption{\label{fig:fig8} One-parameter bifurcation diagrams for the (a) classical and (b) quantum systems with the sine modulation, $N=500$. The maximal element for each value of $U$ is normalized to 1.}
\end{figure}

\begin{figure}
	\includegraphics[width=0.5\textwidth]{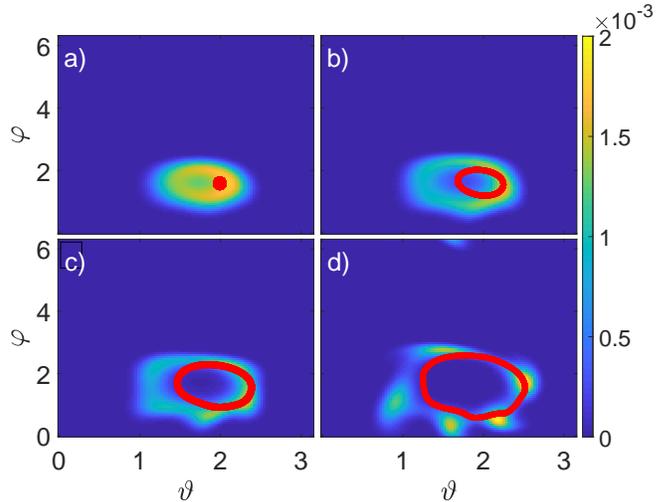}
	\caption{\label{fig:fig9} Husimi distribution for the density matrix after propagation over $100T$ and the stroboscopic Poincar\'{e} map for the mean-field model (red) for different interaction strength: (a) $U=0.1$; (b) $U=0.1125$; (c) $U=0.125$; (d) $U=0.15$. Here $N=500$.}
\end{figure}

We first investigate the nonlinear mean-field equations, Eq.(\ref{eq:3_3}), treating interaction strength $U$ as a bifurcation parameter. There one observes the emergence of an invariant curve for a stroboscopic map $\mathcal{F}~:~\{\theta(mT),\varphi(mT)\}\rightarrow \{\theta((m+1)T),\varphi((m+1)T)\}$ by the Neimark-Sacker bifurcation at $U\sim0.11$ and its succession by period-6 cycle at $U\sim0.18$ (Figs.\ref{fig:fig8}(a), \ref{fig:fig9}). Further it develops into a chaotic attractor, that ultimately disappears through a crisis, when the stable fixed point is recovered.

We interrogate whether quantum equations manifest an analogue of the classical Neimark-Sacker bifurcation and study its properties. 

A sufficiently large number of particles, $N=500$, should bring the system close to the classical limit. The one-parameter bifurcation diagram displays the probabilities to find a given number of particles on the left site of a dimer, taken at stroboscopic times $mT$, $m=0, 1, 2, \ldots$, numerically given by the diagonal elements of the density matrix, $\rho_{n,n}(mT)$.  The qualitative structure of the bifurcation diagram obtained from the mean-field model is reproduced in the quantum case quite well, as for the birth of a torus, as for the onset of chaos and ultimate recovery of a stable fixed point, cf. Fig. \ref{fig:fig8}(b). However, even for $N=500$ some details differ, like the quantum bifurcation within $U\in[0.6,0.7]$, that does not have a counterpart in the mean-field model.

\begin{figure}
	\includegraphics[width=0.5\textwidth]{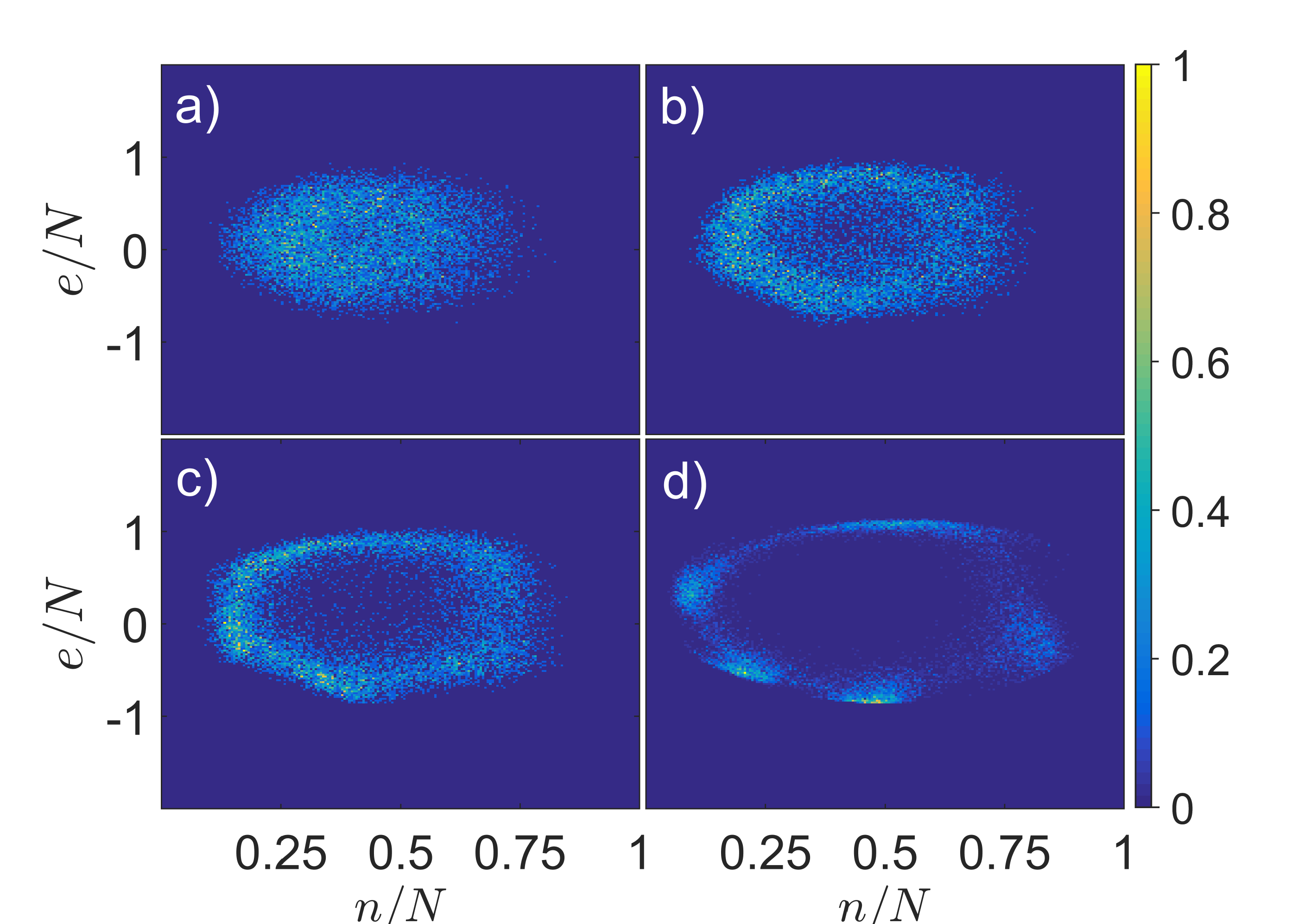}
	\caption{\label{fig:fig9a} Histograms of expectation values for quantum trajectories at stroboscopic times, scaled by the particle number, $n(mT)/N, e(mT)/N$, after propagation over $100T$ for different interaction strength: (a) $U=0.1$; (b) $U=0.1125$; (c) $U=0.125$; (d) $U=0.15$. Here $N=500$.}
\end{figure}

Next, we demonstrate the correspondence between the asymptotic states at stroboscopic Poincar\'{e} sections (for mean-field model) on the plane $\{\vartheta, \varphi\}$ and Husimi distributions for the quantum system, Fig. \ref{fig:fig9}. Indeed, the projection of the quantum attractor onto the classical phase space reveals the emergence of a bagel-shaped distribution after the invariant curve in the Poincar\'{e} section of the mean-field equations. Moreover, one observes the formation of a multi-period orbit on the quantum torus later on, cf. Fig. \ref{fig:fig9}(d), the scenario typical of classical systems.

There are, however, quantitative differences: the bagel is already present at $U=0.1$ for the quantum model with $N=500$, cf. Fig. \ref{fig:fig9}(a), while the mean-field model still has a fixed point; the size of the bagel is slightly greater than the invariant curve in the classical case, cf. Fig. \ref{fig:fig9}(b,c); the formation of a periodic orbit in the quantum case occurs at lower $U$, cf. Fig. \ref{fig:fig9}(d).    

\begin{figure}[t!!!!]
	\includegraphics[width=0.5\textwidth]{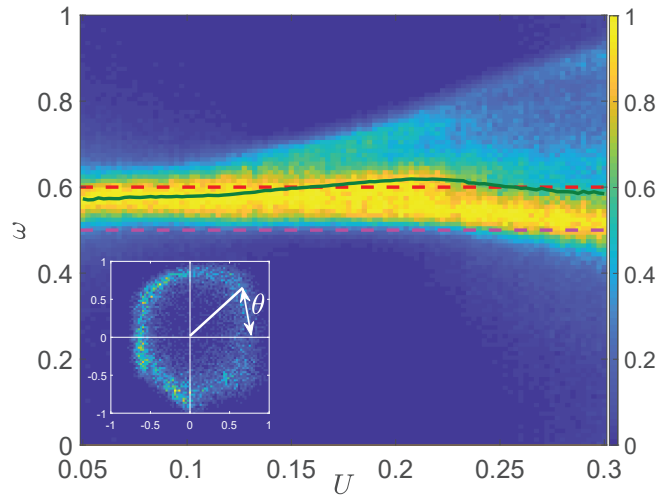}
	\caption{\label{fig:fig12} Color coded histogram of instantaneous rotation number, $\omega$, for quantum trajectories in dependence on interaction strength, $U$. The maximal element for each value of $U$ is normalized to $1$. Solid line corresponds to the average rotation number, dashed lines indicate the levels for rational $\omega=1/2$ and $\omega=3/5$. Here $N=250$.  Inset: the phase $\theta$ is defined as a polar angle for centered and normalized observables. }
\end{figure}

We further investigated, whether quantum Neimark-Sacker bifurcation can be identified directly in quantum observables. Sampling attractor with quantum trajectories (cf. Section \ref{sec:3}), we reconstruct histograms for stroboscopic observables $n(mT),e(mT)$, Eqs.(\ref{eq:5a}), and again witness the emergence of bagel-shaped distributions following quantum bifurcation, cf. Fig.\ref{fig:fig9a} and compare to Fig.\ref{fig:fig9}.

Quantum trajectory unraveling allows to get a deeper insight into the dynamics on quantum torus. That is, one can define the phase $\theta_m$ for each pair of stroboscopic observables,  $n(mT),e(mT)$, as a polar angle, the origin placed at the center of mass of the stroboscopic $2D$ histogram, which is normalized in both dimensions to the same interval $[-1,1]$ (see Fig.\ref{fig:fig12}, inset), and thus calculate the instantaneous rotation (or winding) number 
\begin{equation}
\label{eq:6}
\omega_m=\frac{\theta_m-\theta_{m-1}}{2\pi} \mod 1.
\end{equation}
In classical dynamics, the time-averaged $\omega$ discriminates two cases. For rational $\omega=p/q, \ p,q\in\mathcal{N}$, the invariant torus contains a stable period-$q$ orbit that is observed as an asymptotic solution, while for irrational $\omega$ the trajectories cover the torus densely. The regimes interchange with system parameters.

The dynamics on quantum torus can be characterized by the probability distribution of $\omega_m$, Fig.\ref{fig:fig12}. About the bifurcation point $U^*\approx0.1$ the distribution is well-localized about $\omega=0.58$, different from the period doubling value $1/2$, the bifurcation present for the other parameter values \cite{Ivanchenko2017,Poletti2018}. Our case is analogous to irrational rotation number for classical torus, so that quantum trajectories densely cover the bagel. The average rotation number increases with interaction strength, and becomes rational $\omega=3/5$ at $U\approx0.15$. Then the stroboscopic distributions obtain a clear footprint of the corresponding period-$5$ structure (Figs.\ref{fig:fig9}(c,d), \ref{fig:fig9a}(c,d)), as in the classical case. 

\begin{figure}[t!!!!]
	\includegraphics[width=0.5\textwidth]{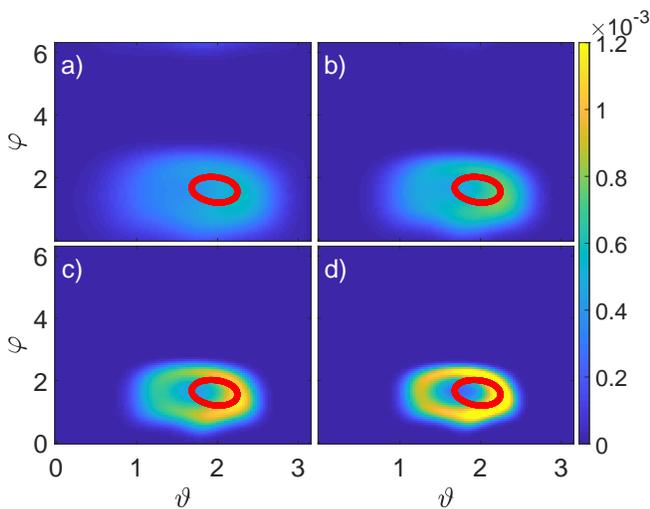}
	\caption{\label{fig:fig10} Husimi distribution for the density matrix after propagation over $100T$ and the stroboscopic Poincar\'{e} map for mean-field model (red) for the different number of particles: (a) $N=50$; (b) $N=100$; (c) $N=250$; (d) $N=500$. Here $U=0.1125$.}
\end{figure}

\begin{figure}[t!!!!]
	\includegraphics[width=0.5\textwidth]{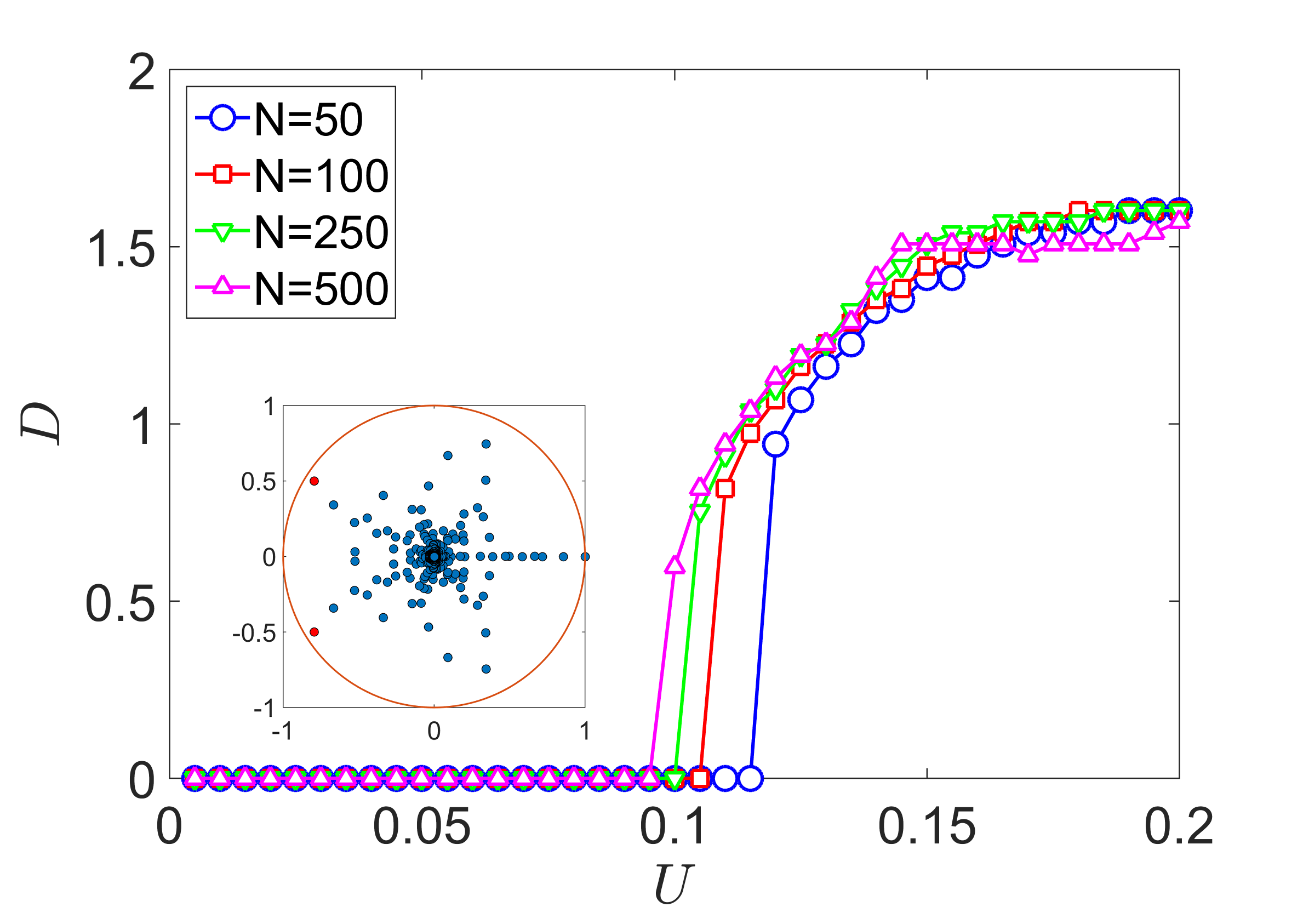}
	\caption{\label{fig:fig11} Diameter of a bagel, $D$, in the Husimi distribution in dependence on the interaction strength, $U$, after propagation over $100T$ for the different number of particles: $N=50$ (blue); $N=100$ (red); $N=250$ (green); $N=500$ (purple). Inset: eigenvalues of Floquet map, $\mathcal{P}_F$, on a complex plane about the quantum bifurcation point (blue) and a unit circle (orange), $U=0.12, N=50$; red points mark $\mu_{2,3}$.}
\end{figure}

The bifurcation also affects the spectral properties of the system. This can be demonstrated by calculating and diagonalizing the Floquet map $\mathcal{P}_F=\mathcal{T}\exp[\int_0^T \mathcal{L}dt]$, where $\mathcal{T}$ is time-ordering operator, that describes evolution of the density operator over a period of modulation under Eq.(\ref{eq:1}) \cite{Hartmann2017}.
The largest eigenvalue of its spectrum, $\{\mu_k\}$, $k=1,\ldots,(N+1)^2$, is always unity, $\mu_1=1$, the rest are inside the unit circle, as per the dissipative nature of $\mathcal{L}$. We follow the next to largest eigenvalues and find a conjugated pair that approaches the unit circle at the point of quantum bifurcation, $\mu_{2,3}\approx e^{\pm i\theta_0}$ (Fig.\ref{fig:fig11}, inset). Its complex phases are consistent with the rotation number, $\theta_0\approx2\pi\omega$. Repeating the treatment of Ref.\cite{Poletti2018}, one obtains that the two-time correlation of an observable in the asymptotic regime is {\it incommensurate} with the time-scale of modulation. Note that the spectral gap $1-|\mu_{2,3}|$ decreases with $N$, yielding the classical conjugate multipliers of the limit cycle at the Neimark-Sacker bifurcation point \cite{Kuznetsov} in the infinite-size limit, $N\rightarrow\infty$. It also gives an estimate of relaxation time from a random initial condition to the asymptotic state,  $t\sim (1-|\mu_{2,3}|)^{-1}$, which can become very large with $N$, it follows.

Finally, we investigate the dependence on the number of bosons, $N$. The classical limit is formally obtained for $N\rightarrow\infty$, but the traits of classical Neimark-Sacker bifurcation are reproduced deep in the quantum regime, as soon as $N\sim25$. In fact, the number of particles, $N$, can be considered as a bifurcation parameter itself. For instance, one can explore the case $U=0.1125$, when an invariant curve is already present in the mean-field model, cf. Figs.\ref{fig:fig9}(b), \ref{fig:fig9a}(b). However, the Husimi distribution for the quantum dimer with the relatively small number of particles, $N=50$, is still unimodal, cf. Fig.\ref{fig:fig10}(a). Increasing $N$ one observes a transformation to a bagel shape (Fig.\ref{fig:fig10}(b-d)), i.e. Neimark-Sacker bifurcation with the system size $N$.

In classical nonlinear systems the size of the invariant curve generically scales as a square root above the bifurcation point \cite{Kuznetsov}. For the quantum bifurcation we define the diameter of the bagel $D$ in the Husimi distribution as the distance between two maxima in the section $\phi=\pi/2$. When the distribution is unimodal, we set $D=0$. The resulting curves $D(U)$ confirm a pronounced dependence on the number of particles, in particular, for the bifurcation point, cf. Fig. \ref{fig:fig11}. The scaling of $D(U)$ about the bifurcation point also depends on the number of bosons and is not universal.

\section{Conclusions}\label{sec:8}

We found and investigated the quantum counterpart of the classical Neimark-Sacker bifurcation in an open periodically modulated quantum dimer. The classical bifurcation in a dissipative nonlinear system is a birth of torus from a limit cycle (or an invariant curve from a fixed point for a time-discrete map). Concurrently, a conjugate pair of complex Floquet multipliers of a periodic trajectory crosses the unit circle. 

The quantum Neimark-Sacker bifurcation is manifested by emergence of bagel-shaped stroboscopic distributions, as for Husimi projection on the classical phase space, as for quantum observables. The bifurcation is also seen in the spectral properties of the corresponding Floquet map, as a pair of its conjugate eigenvalues approaching the unit circle.  Quantum trajectories technique allows for unraveling the dynamics on the ``quantum torus'', and a suitably generalized rotation number on it. In analogy to the classical case, irrational and rational rotation numbers yield bagel (dense torus) and multi-modal (periodic orbit on torus) distributions. The quantum bifurcation depends on the system size -- the number of bosons, and more sensitively for fewer particles, away from the mean-field limit. As such the number of particles is a bifurcation parameter.

We expect that the bifurcation can be found in the other open non-equilibrium quantum setups, to name a modulated open Dicke model, where its classical counterpart has been reported \cite{gong2018}. Thereby, cavity and circuit QED systems appear the first candidates for its experimental observation.  

\section{Acknowledgments}\label{acknowledgment}

The authors acknowledge support of the President of Russian Federation grant No. MD-6653.2018.2 and Russian Foundation for Basic Research grant No. 18-32-20221.

\end{document}